\documentclass[a4paper,11pt]{article}

\usepackage{jheppub} 

\usepackage[T1]{fontenc} 
\usepackage{comment}

\title{\boldmath  Fluid description of gravity on a timelike cut-off surface: beyond Navier-Stokes equation}

 \author{Shounak De {and}}
 \author{Bibhas Ranjan Majhi}
 \affiliation{Department of Physics, Indian Institute of Technology Guwahati, Guwahati 781039, Assam, India}

\emailAdd{shounakde@iitg.ac.in}
\emailAdd{bibhas.majhi@iitg.ac.in}

\abstract{Over the past few decades, a host of theoretical evidence has surfaced that suggest a connection between theories of gravity and the Navier-Stokes (NS) equation of fluid dynamics. It emerges out that a theory of gravity can be treated as some kind of fluid on a particular surface. Motivated by the work carried out by Bredberg et al (JHEP \textbf{1207}, 146 (2012)) \cite{Bredberg:2011jq}, our paper focuses on including certain modes to the vacuum solution which are consistent with the so called hydrodynamic scaling and discuss the consequences, one of which appear in the form of Damour Navier Stokes (DNS) equation with the incompressibility condition. We also present an alternative route to the results by considering the metric as a perturbative expansion in the hydrodynamic scaling parameter $\epsilon$ and with a specific gauge choice, thus modifying the metric. It is observed that the inclusion of certain modes in the metric corresponds to the solution of Einstein's equations in presence of a particular type of matter in the spacetime. This analysis reveals that gravity has both the NS and DNS description not only on a null surface, but also on a timelike surface. So far we are aware of, this analysis is the first attempt to illuminate the possibility of presenting the gravity dual of DNS equation on a timelike surface. In addition, an equivalence between the hydrodynamic expansion and the near-horizon expansion has also been studied in the present context.}

\begin{document} 
\maketitle
\flushbottom

\section{Introduction}
\label{sec:intro}

The Einstein's field equations of gravity and the Navier-Stokes (NS) equation governing the dynamics of fluids are two of the most important differential equations in physics and mathematics. While the Einstein equation universally governs the long-distance behaviour of essentially any gravitating system, the NS equation governs the hydrodynamic limit of essentially any fluid. Any connection between these two rich non-linear structures is bound to be interesting and hints of such a connection have surfaced in various forms over the last four decades.
The connections established between gravity and fluid dynamics lends credence to the notion that gravity is not a truly fundamental phenomenon, but an emergent one. Apart from fluid-gravity duality, a host of theoretical evidence \cite{Padmanabhan:2009vy} suggest gravity could be an emergent phenomenon with the gravitational field equations having the same status as the equations of gas dynamics.

One of the earliest works in this regard appears in the doctoral thesis of Damour \cite{Damour:1979wya}, wherein there are suggestions of a relation between horizon and fluid dynamics. This work shows that Einstein's field equations when projected onto a null surface, exhibits a mathematical structure very similar to NS equation but with an additional non-linear term, now known as the Damour-Navier-Stokes (DNS) equation. The DNS equation governs the geometric data on any null surface.
Such a connection has also been obtained in the membrane paradigm approach by Price and Thorne in \cite{Price:1986yy}. Due to the presence of an additional term, they refer to the fluid dynamical equation as Hajicek equation instead of the usual NS equation. A corresponding action formulation has been discussed extensively by Parikh and Wilczek in \cite{Parikh:1997ma}. The same has also been done in \cite{Kolekar:2011gw} for DNS.

In a more recent work by Bredberg et al \cite{Bredberg:2011jq}, which forms the basis of our work, a general solution of the vacuum Einstein equations with certain prescribed boundary data is presented in ($p+2$) dimensions ($p$ corresponds to all angular coordinates while $2$ stands for one time and one radial coordinate) using the well-known hydrodynamic expansion of the incompressible NS equation in ($p+1$) dimensions (here $p$ is, like earlier, collection of all angular coordinates and 1 refers to only the timelike coordinate). In this paper, they seek a relation between the ($p+ 2$)-dimensional Einstein and ($p + 1$)-dimensional NS equation by the construction of a metric in accordance with a scaling symmetry the incompressible NS equation possess. The metric construction is an expansion in the hydrodynamic scaling parameter $\epsilon$ and is parameterized by the velocity field $v_i(x^i,\tau)$ and pressure $P(x^i,\tau)$ that solves the full nonlinear incompressible NS equation. It may be worth mentioning that the metric in  \cite{Bredberg:2011jq} is constructed so as to strictly obtain incompressible NS dynamics by conserving the corresponding Brown-York stress tensor on a timelike hypersurface $r = r_c$, where $r_c$ lies outside the black hole horizon $r_h$ (also, an earlier seminal work \cite{Bhattacharyya:2008kq} is done in the AdS/CFT context). The choice of the metric thus lies on the following facts. It has been obtained perturbatively in the hydrodynamic expansion parameter around Rindler space-time and up to certain gauge conditions and reparameterizations, satisfies certain boundary conditions and solves the vacuum Einstein's equations. The explicit meaning of these statements will be made clear in the next paragraph. 
This cut-off surface approach has been applied in various cases \cite{Huang:2011he}. For example, it was extended for higher curvature gravity theories \cite{Chirco:2011ex, Bai:2012ci, Cai:2012mg, Zou:2013ix, Hu:2013lua} as well as for the AdS \cite{Cai:2011xv, Huang:2011kj} and dS \cite{Anninos:2011zn} gravity theories (for other theories, like black branes, see \cite{Ling:2013kua}). Also a corresponding relativistic situation has been discussed extensively in \cite{Eling:2012ni}. Symmetries of the vacuum Einstein equations have been exploited to develop a formalism for solution generating transformations of the corresponding NS fluid duals in \cite{Berkeley:2012kz}. The fluid description on the Kerr horizon has also been explored extensively in \cite{Lysov:2017cmc} (see \cite{Wu:2013aov} for the isolated horizon case).

Our motivation is to analyse the metric construction more deeply and see how far we can generalise the form of the metric which abide by the same scaling laws and thereby investigate the corresponding consequences. On analysing the vacuum metric we realized that certain additional modes can still be incorporated in the metric and in accordance with the relevant hydrodynamic scaling. For example, terms like $(\partial_i v_j + \partial_j v_i) dx^i dx^j$ and $v^2 \delta_{ij}dx^i dx^j$ are perfectly allowed modes of the order of ${\epsilon^2}$ and hence can be added to the metric of \cite{Bredberg:2011jq}. In our work we try to figure out the implications of retaining these scaling consistent modes which serve as modifications to the original vacuum solution presented in \cite{Bredberg:2011jq}. One such consequence due to the presence of the term $\partial{_{_(i}}v{_{j_)}}$ is that the modified metric generates fluid dynamics described by the equation which contains an additional term proportional to $\partial_i v^2$, while retaining the incompressibility condition. Interestingly, such a structure is similar to that obtained in \cite{Padmanabhan:2010rp} where the DNS equation is expressed in terms of coordinates adapted to a null surface. As discussed in length in the work by Padmanabhan \cite{Padmanabhan:2010rp}, the key structural difference of the DNS equation in comparison to the NS equation is the presence of a $\partial_i v^2$ kind of term. We obtain a seemingly close structure which has only the kinematic viscosity (thereby calling it a "restricted" DNS equation) when one keeps a term of the type $(\partial_iv_j+\partial_jv_i) dx^i dx^j$ in the original metric and approaching in the same route as in \cite{Bredberg:2011jq}. We show later in the subsequent sections that these additional mode(s) result from defining an appropriate bulk matter and the corresponding generalized metric solves the Einstein field equations with an appropriate energy-momentum tensor.

This is a new scenario quite in contrary to the approach by Damour \cite{Damour:1979wya} which was set in context to null surface dynamics alone, whereas the present one is constructed on a timelike surface. Such an investigation fulfils an important purpose. In literature, the NS equation has both types of gravity duality: one on the timelike surface and another on the null surface. On the contrary, so far we are aware of, the DNS equation has a gravity dual constructed on a null surface. So the present analysis can be regarded as the first attempt to have a gravity description of the DNS equation on a timelike surface, along with a particular constraint. 

We also describe an alternative framework in which the geometry described in  \cite{Bredberg:2011jq}, is now viewed as a perturbative expansion of the leading order metric and leading to NS or DNS dynamics depending on the manner in which the metric is modified at a particular order. In this case, the extrinsic curvature and the surface stress-tensor is also taken to have a linear order correction. The perturbative expansion follows a specific choice of fixed gauge conditions. This gauge condition confirms the invariance of the structure of the seed metric on the cut-off surface. In both the frameworks, the equivalence between the hydrodynamic expansion and the near-horizon expansion approaches is explored. Implications of these results are finally discussed at the end.

The organization of the paper is as follows. In Section \ref{scale}, we briefly review the well known scaling symmetry of the incompressible NS equation which serves as the basis of our main goal. In the next Section, we present a generalized metric where the additional scaling consistent tensor modes are sourced by an appropriately defined bulk matter characterized by an energy-momentum tensor and the metric is shown to solve the complete Einstein field equations, i.e., in presence of matter, upto the order the bulk solution is presented. The constraint equations on the cut-off boundary surface is shown to reduce to the incompressible DNS-like fluid equations. In Section \ref{pert}, we present an alternative framework to represent the gravity dual of this constrained DNS fluid, with the metric now modified by a specific choice of gauge conditions and the extrinsic curvature along with the Brown-York stress tensor taken to have linear order corrections only. The equivalence between the hydrodynamic expansion and the near-horizon expansion approaches, similar in spirit to \cite{Bredberg:2011jq} is explored in Section \ref{near}. Next, we present some discussions that could possibly illuminate our understanding of certain key aspects.  Finally, we present our conclusions in Section \ref{conc}. We present the detailed proof of the scale invariance of the restricted DNS-like equation in Appendix \ref{App1}. In Appendix \ref{App2}, we show the explicit computation of the Ricci tensor components for the metrics presented in the paper, but whose coefficients have been taken to be pre-factored with suitable multiplicative constants, indicating as to how various modes contribute to the Ricci tensor at various orders. 

The {\it notations} used throughout the paper are clarified as follows: All uppercase Latin letters (i.e., $A, B, C,$ etc.) denote the bulk spacetime co-ordinate indices while the lowercase Latin letters (i.e., $a, b, c$ etc.) are used only for denoting the transverse coordinates. The Greek letters (i.e., $\alpha, \mu, \nu$ etc.) denote the coordinates on the boundary cutoff hypersurface $\Sigma_c$.

\section{\label{scale}Setup: Scaling laws} 
In this section we shall briefly review the scaling symmetry of the incompressible NS equation, which forms the basis of the work \cite{Bredberg:2011jq} and also lays the foundation of our analysis. Now, if the amplitudes of the solution space $(v_i, P)$ of the incompressible NS equation is scaled down by the parameter $\epsilon$:
 \begin{align}
 v_i^\epsilon(x^i,\tau)&=\epsilon v_i(\epsilon x^i,\epsilon^2 \tau)~;
 \nonumber 
 \\
 P^\epsilon(x^i,\tau)&=\epsilon^{2}P(\epsilon x^i,\epsilon^2 \tau)~,
 \label{set1}
 \end{align}
then the NS equation remains invariant under the above scaling transformation, thus generating a family of solutions parameterized by $\epsilon$ from the original solution space. Any deviations from ideality lead to certain typical corrections which vanish under the scaling law (\ref{set1}), as shown in \cite{Bredberg:2011jq}. A term proportional to $\partial_i v^2$ arises in our analysis later in the paper, as an addition to the incompressible NS equation. This resulting equation is also shown to obey the same scaling laws (\ref{set1}) (the explicit proof is shown in Appendix \ref{App1}). The hydrodynamic scaling parameter $\epsilon$ serves as the expansion parameter for the metric in \cite{Bredberg:2011jq}. The hydrodynamic scaling of the spatial and time derivatives along with the pair $(v_i,P)$ follows as:
 \begin{align}
 v^i \sim \mathcal{O} (\epsilon)\,, \quad  P \sim \mathcal{O} (\epsilon^2)\,, \quad \partial_i  \sim \mathcal{O} (\epsilon)\,,\quad \partial_{\tau} \sim \mathcal{O} (\epsilon^2)~.
 \label{set2}
 \end{align}
In our analysis in the subsequent sections, we focus on including certain additional modes consistent with the above hydrodynamic scaling sourced by an appropriately defined bulk matter tensor and working with a modified metric than in \cite{Bredberg:2011jq}. We then strive to discuss the consequences of this generalized metric with these scaling consistent modes in a formalism similar to \cite{Bredberg:2011jq} and also discuss it in a different framework. We shall see that it not only includes the NS equation, but also gives birth to a {\it restricted} type of DNS equation which also preserves its structure under the scaling (\ref{set1}) of the fluid parameters.

\section{\label{dns}Metric and incompressible DNS}
We present a metric keeping some more possible terms consistent with the scaling argument as discussed above. The leading order base metric is taken to be flat and is in (ingoing) Rindler form in Eddington-Finkelstein coordinates. The metric retains its original structure as in \cite{Bredberg:2011jq}, albeit for the two additional scaling consistent terms that now appear at $\mathcal{O}(\epsilon^2)$ which are pre-factored with constants $\tilde{a_1}$ and $\tilde{a_2}$. Of course, these terms vanish on the timelike hypersurface $r=r_c$ as well, such that the induced metric is flat. We therefore present a metric of the following form: 

\begin{align}
ds_{p+2}^2= g_{A B} dx^A dx^B &= -rd\tau^2 + 2d\tau dr + dx_idx^i  \nonumber \\
&-2\bigg(1-\frac{r}{r_c}\bigg)v_i dx^i d\tau - 2\frac{v_i}{r_c} dx^i dr \nonumber \\
&+\bigg(1-\frac{r}{r_c}\bigg)\bigg[(v^2+2P)d\tau^2 + \frac{v_i v_j}{r_c}dx^i dx^j + \tilde{a_1}\frac{v^2 \delta_{ij}}{r_c}dx^i dx^j \nonumber \\
&+ \tilde{a_2}(\partial_i v_j + \partial_j v_i) dx^i dx^j\bigg] + \bigg(\frac{v^2}{r_c} + \frac{2P}{r_c}\bigg)d\tau dr \nonumber \\
&-\frac{r^2-r_c^2}{r_c} \partial^2 v_i dx^i d\tau + \dots
\label{1.01} 
\end{align}
where $v_i = v_i(x^i, \tau)$ and $P(x^i, \tau)$ are independent of the radial coordinate $r$. The first line is the flat base metric of $\mathcal{O}(\epsilon^0)$ while the second line contains all possible $\mathcal{O}(\epsilon)$ terms. The next terms that follow in the third and fourth lines are of $\mathcal{O}(\epsilon^2)$. 
The metric presented in \cite{Bredberg:2011jq}, i.e., (\ref{1.01}) with $\tilde{a_1}$ and $\tilde{a_2}$ set identically to zero, is a solution of the dynamical vacuum Einstein field equations up to $\mathcal{O}(\epsilon^3)$ while the constraint equations generate incompressible NS fluid dynamics on the cutoff surface $r = r_c$. Thus, as it stands, the incorporation of the terms with coefficients $\tilde{a_1}$ and $\tilde{a_2}$ ceases the metric (\ref{1.01}) to be a solution of the vacuum field equations, leading to corrections in the Ricci tensor at $\mathcal{O}(\epsilon^2)$. This can be checked by a direct computation of the components of the Ricci tensor. The non zero contributions at $\mathcal{O}(\epsilon^2)$ to the Ricci tensor $R_{A B}$ are as follows (an extensive, general computation of the Ricci tensor components is presented in Appendix \ref{App2}):
\begin{align}
R^{(2)}_{r \tau} &= - \frac{1}{r} R^{(2)}_{\tau \tau} = \frac{\tilde{a_1}}{4 r_c^2} p v^2 \,, \label{1.02a} \\
R^{(2)}_{i j} &= \frac{\tilde{a_1}}{2 r_c^2} v^2 \delta_{i j} + \frac{\tilde{a_2}}{2 r_c} (\partial_i v_j + \partial_j v_i) \,, \label{1.02b}
\end{align}
where $p = \delta_{i i}$ refers to the number of angular dimensions. The relations (\ref{1.02a}) and (\ref{1.02b}) are also in accordance to the corrections predicted in \cite{Compere:2011dx} (refer to Eq. (4.2) of \cite{Compere:2011dx}).

In the cut-off surface approach to the fluid/gravity correspondence, one first constructs a metric by using the perturbation technique where the hydrodynamic expansion parameter $\epsilon$ acts as the perturbative parameter; such that it solves the Einstein field equations. Then one concentrates on the constraint part of the Einstein field equations on the cut-off boundary surface, on which the initial data is prescribed. This gives the fluid dynamical equations at different orders in $\epsilon$. More precisely, in the Arnowit-Deser-Misner (ADM) formalism, the Einstein equations lead to two class of equations. The initial data is specified by the induced metric $\gamma_{\mu \nu}$ on the cutoff boundary surface $\Sigma_c$. In this construct, the extrinsic curvature $K_{\mu \nu}$ (which are analogous to the canonical momenta), also needs to be specified on $\Sigma_c$ to completely prescribe the initial data on this cutoff surface. These are all determined by the constraint Einstein equations. One of them is given by \cite{Gourgoulhon:2007ue}
\begin{equation}
D_\mu T_{\textrm{(BY)}}^{\mu \nu} = \kappa\gamma^{\nu B} N^A T_{A B}
\label{JHEP1}
\end{equation}
where
\begin{align}
T^{\, \textrm{(BY)}}_{\mu \nu} = (\gamma_{\mu \nu} K - K_{\mu \nu})~,
\label{1.02}
\end{align}
 is the  Brown-York stress tensor and $T_{AB}$ is the matter stress-tensor. In the above, $\gamma_{\mu \nu}=g_{\mu\nu}-N_\mu N_\nu$ is the induced metric on the $r=r_c$ cutoff hypersurface $\Sigma_{c}$ and $\kappa=8\pi G$. $N^A$ is the unit normal to the induced surface satisfying $N_AN^A=1$. The extrinsic curvature $K_{\mu\nu}$ is defined to be the Lie derivative of the induced metric along the normal $N^{\mu}$ on the hypersurface:
\begin{align}
K_{\mu \nu} = \frac{1}{2} \pounds_{N} \gamma_{\mu \nu}~,
\label{1.03}
\end{align}
and $K$ is the trace of the above quantity; i.e. $K=\gamma^{\mu\nu}K_{\mu\nu}$. The other constraint equation is \cite{Gourgoulhon:2007ue}
\begin{equation}
^{(p+1)}R+K^2-K_{\mu\nu}K^{\mu\nu} = 2\kappa T_{AB}N^AN^B~.
\label{JHEP2}
\end{equation}
In the above relation, $^{(p+1)}R$ is the Ricci scalar defined on the induced surface.
We shall see later that in our present analysis the right hand side is $\mathcal{O}(\epsilon^4)$.

We have already seen that the metric (\ref{1.01}) is not a solution to the vacuum Einstein equations upto $\epsilon^3$ order as there are non vanishing contributions to the Ricci tensor $R_{A B}$ at $\mathcal{O}(\epsilon^2)$ itself, given by (\ref{1.02a}) and (\ref{1.02b}). Therefore, we now aim to support our metric (\ref{1.01}) by establishing it as a solution to the Einstein field equations in presence of matter (unlike the solution in \cite{Bredberg:2011jq} which solves the vacuum field equations):
\begin{align}
G_{A B} = R_{A B} - \frac{1}{2} R g_{A B} = \kappa T_{A B}~,
\label{1.022a}
\end{align}
where $\kappa=8\pi G$.
We argue the fact that the correction term(s) in (\ref{1.01}) are in fact sourced by an appropriately defined bulk energy-momentum tensor $T_{A B}$ characterizing the matter. The only constraint we impose on this energy-momentum tensor $T_{A B}$ is for it to be traceless, i.e., 
\begin{align}
T = g^{A B} T_{A B} = 0 \,,
\label{1.02c} 
\end{align}
upto $\mathcal{O}(\epsilon^2)$, where $T$ is the trace of the symmetric energy-momentum tensor. We shall show that the imposition of the traceless condition on the matter energy-momentum tensor is not arbitrary but instead lead to important fluid dynamical consequences. The traceless condition on $T_{A B}$ implies an identically vanishing Ricci scalar, i.e., $R = 0$, and the Einstein field equations (with matter) (\ref{1.022a}) are thereby reduced to the form:
\begin{align}
R_{A B} = \kappa T_{A B} \,.
\label{1.02d}
\end{align}
The components of the matter stress-tensor can be determined by the non-vanishing components of $R_{AB}$ and the above form of Einstein equations. 
In view of the relations (\ref{1.02a}) and (\ref{1.02b}), it is evident from (\ref{1.02d}) that the non zero components of the energy-momentum tensor $T_{A B}$ up to $\mathcal{O}(\epsilon^2)$ are:
\begin{align}
\kappa T^{(2)}_{\tau r} = R^{(2)}_{r \tau} &=  \frac{\tilde{a_1}}{4 r_c^2} p v^2 \,, \label{1.021} \\
\kappa T^{(2)}_{\tau \tau} = R^{(2)}_{\tau \tau} &= - \frac{\tilde{a_1}}{4 r_c^2} p r v^2 \,, \label{1.022} \\
\kappa T^{(2)}_{i j} = R^{(2)}_{i j} &= \frac{\tilde{a_1}}{2 r_c^2} v^2 \delta_{i j} + \frac{\tilde{a_2}}{2 r_c} (\partial_i v_j + \partial_j v_i) \,. \label{1.023}
\end{align}
With the above identification of the matter stress-tensor, let us now concentrate on the constraint Einstein equations (\ref{JHEP1}).
We start with the calculation of the unit normal on the timelike hypersurface $r = r_c$ for the metric (\ref{1.01}). It is given by $N_A=A_0\partial_{A}r$ where the normalization $A_0$ is determined by the condition $g^{A B} N_A N_B = 1$. On $r=r_c$ it turns out to be:
\begin{align}
N^{A}\partial_{A} = \frac{1}{\sqrt{r_c}} \partial_{\tau} + \sqrt{r_c} \bigg(1-\frac{P}{r_c}\bigg) \partial_r + \frac{v^i}{\sqrt{r_c}} \partial_i + \mathcal{O}(\epsilon^3)~,
\label{1.04}
\end{align}
which remains identical to the unit normal derived in \cite{Bredberg:2011jq} up to the required orders in $\epsilon$, although the metrics are not completely identical.
With the above definition (\ref{1.02}) for the Brown-York stress tensor and the expression for unit normal (\ref{1.04}), the components of the stress-tensor of the metric (\ref{1.01}) on $\Sigma_c$ are derived to be
\begin{align}
T^{\, \textrm{(BY)}}_{\mu \nu}dx^{\mu}dx^{\nu} &= \,\frac{1}{2} \bigg[ \frac{dx^2_i}{\sqrt{r_c}} + \frac{v^2}{\sqrt{r_c}} d\tau^2 - 2\frac{v_i}{\sqrt{r_c}} dx^i d\tau +\frac{(v_iv_j + P\delta_{ij} + \tilde{a_1} v^2 \delta_{ij})}{r^{3/2}_c} \nonumber \\
&+\frac{(\tilde{a_2}-2) \partial_i v_j + \tilde{a_2} \partial_j v_i}{\sqrt{r_c}} dx^i dx^j \bigg]+ \mathcal{O}(\epsilon^3)~.
\label{1.05} 
\end{align}
We note the appearance of the additional terms with coefficients $\tilde{a_1}$ and $\tilde{a_2}$ in comparison to the expression derived in \cite{Bredberg:2011jq}. 

We now consider the constraint equations (\ref{JHEP1}) that are to be satisfied on $\Sigma_c$, which constrain our initial data through the equations:
\begin{align}
D_{\mu} T^{\mu \nu}_{\textrm{(BY)}} \bigg{|}_{\Sigma_c} &= \kappa \gamma^{\nu B} N^A T_{A B} = \gamma^{\nu B} N^A R^{(2)}_{A B} \,,
\label{1.06}
\end{align}
where we have used the Einstein field equations (\ref{1.02d}) for the last equality. 
With the free index $\nu$ assuming the time co-ordinate $\tau$, we arrive at the first nontrivial equation at $\mathcal{O}(\epsilon^2)$:
\begin{align}
\partial_i v^i = 0 \,,
\label{1.07}
\end{align}
which is the incompressibility condition satisfied by the velocity field $v^i$ of the fluid. 
Interestingly, the incompressibilty condition is not affected by the new terms appearing in the metric (\ref{1.01}). This is due to the fact that it is solely derived from the $\mathcal{O}(\epsilon)$ terms in the metric.
With the other choice $\nu=i$, we obtain at $\mathcal{O}(\epsilon^3)$: 
\begin{eqnarray}
\partial_{\tau} v_i + v^j \partial_j v_i + \tilde{a_1} \partial_i v^2 + \partial_i P - \eta_{\textrm{eff}} \partial^2 v_i = \frac{\tilde{a_1}}{r_c} v_i v^2 + \tilde{a_2} (v^j \partial_i v_j + v^j \partial_j v_i) \,,
\label{1.08}
\end{eqnarray}
where $\eta_{\textrm{eff}} = \frac{2 - \tilde{a_2}}{2} \eta$ with $\eta = r_c$. 
The result (\ref{1.08}) can be interpreted as two different scenarios depending on the contributions from the modes with coefficients $\tilde{a_1}$ and $\tilde{a_2}$.

For $\tilde{a_1} = 0$ and $\tilde{a_2} = 0$, Eq. (\ref{1.08}) reduces to 
\begin{align}
\partial_{\tau} v_i + v^j \partial_j v_i + \partial_i P - \eta \partial^2 v_i = 0~.
\label{1.09}
\end{align}
which is the conventional Navier-Stokes dynamics with kinematic viscosity $\eta_\textrm{eff}=\eta=r_c$, this situation being what was exactly tackled in \cite{Bredberg:2011jq}. As seen from relations (\ref{1.021})-(\ref{1.023}) and (\ref{1.02a})-(\ref{1.02b}), corrections to the Ricci tensor and components of the matter energy-momentum tensor vanish and consequently the appropriate metric solves the vacuum Einstein field equations up to $\mathcal{O}(\epsilon^3)$. 

However, with $\tilde{a_1} = 0$ and considering contributions from the tensor mode $(\partial_i v_j + \partial_j v_i) dx^i dx^j$ solely, i.e. with $\tilde{a_2} \neq 0$, the non zero corrections to the Ricci tensor reduces to 
\begin{align}
R^{(2)}_{i j} &= \frac{\tilde{a_2}}{2 r_c} (\partial_i v_j + \partial_j v_i) \,, \label{1.08b} 
\end{align}
which we clearly see from (\ref{1.02a}) and (\ref{1.02b}). Consequently, our matter energy-momentum tensor $T_{A B}$ has non zero contributions of the form:
\begin{align}
\kappa T^{(2)}_{i j} = R^{(2)}_{i j} &= \frac{\tilde{a_2}}{2 r_c} (\partial_i v_j + \partial_j v_i) \,. \label{1.08c}
\end{align}
which again follows from (\ref{1.021})-(\ref{1.023}). With these conditions, our fluid dynamical equation (\ref{1.08}) gives at $\mathcal{O}(\epsilon^3)$:
\begin{align}
\partial_{\tau} v_i + (1 - \tilde{a_2}) v^j \partial_j v_i - \bigg(\frac{\tilde{a_2}}{2}\bigg) \partial_i v^2 + \partial_i P - \eta_{\textrm{eff}} \partial^2 v_i = 0~.
\label{1.10a}
\end{align}
Explicitly setting $\tilde{a_2} = - 1$, we get from (\ref{1.10a}):
\begin{align}
\partial_{\tau} v_i + 2 v^j \partial_j v_i + \frac{1}{2} \partial_i v^2 + \partial_i P - \eta_{\textrm{eff}} \partial^2 v_i = 0~.
\label{1.101}
\end{align}
Redefining our time co-ordinate, we make the transformation $\partial_{\tau} + v^j \partial_j \to \partial_{\tau}$ and (\ref{1.101}) reduces to:
\begin{align}
\partial_{\tau} v_i +  v^j \partial_j v_i + \frac{1}{2} \partial_i v^2 + \partial_i P - \eta_{\textrm{eff}} \partial^2 v_i = 0~.
\label{1.10}
\end{align}
We note the presence of the third term which clearly isolates it from the conventional NS equation. Below we shall argue that such a form of equation was obtained earlier by Damour \cite{Damour:1979wya}, now known as Damour-Navier-Stokes (DNS) equation, in a different context.

Before delving into the analysis of equation (\ref{1.10}), we try to understand what the traceless condition imposed on the bulk matter $T_{A B}$ means from a purely fluid perspective. To this end, we have using (\ref{1.02c}) and (\ref{1.08c}):
\begin{align}
T = g^{A B \, (0)} T_{A B}^{(2)} = \frac{\tilde{a_2}}{2 r_c \kappa} \delta^{i j} (\partial_i v_j + \partial_j v_i) = 0 \implies \partial_i v^i = 0 \quad (\because \tilde{a_2} \neq 0 )\,.
\end{align}
Thus, we see that a traceless bulk matter described by (\ref{1.08c}) ultimately reduces to the incompressibility condition, independently of how we initially arrived through the constraint equation at $\mathcal{O}(\epsilon^2)$ given by (\ref{1.07}). This result gives considerable backing to the traceless condition we imposed as an ansatz on the bulk matter $T_{A B}$ and is not arbitrary in any sense. Also it may be worthwhile to mention that the other constraint equation (\ref{JHEP2}), which is the Hamiltonian constraint, is automatically satisfied. This is understood in the following way. The only non-vanishing component of matter stress-tensor in this case is $T_{ij}$ which is $\mathcal{O}(\epsilon^2)$ and  $N^i$ is $\epsilon$ order. Therefore the right hand side of (\ref{JHEP2}) is $\epsilon^4$ order.

It has been shown that the Einstein field equations when projected onto a null surface leads to an equation that bears striking resemblance to the form obtained in (\ref{1.10}). This equation has a purely geometrical interpretation. Later Padmanabhan showed in \cite{Padmanabhan:2010rp} that, the Einstein's equation when projected onto a null surface, represented by a metric adapted to null coordinates, leads to a more convenient form of the DNS equation. Our equation (\ref{1.10}) closely resembles his form. If one compares (\ref{1.10}) with Eq. (30) of \cite{Padmanabhan:2010rp}, it is seen that in our analysis we obtain a restricted version of the full DNS equation. The full DNS equation (30) in \cite{Padmanabhan:2010rp} has additional terms to what is obtained in (\ref{1.10}). This is because we have an additional constraint in the form of the incompressibility condition (\ref{1.07}) which when imposed on the full DNS equation leads to an incompressible DNS equation (\ref{1.10}). As a result, we are left with only the kinematic viscosity $\eta$ (which is characteristic of an incompressible fluid), in contrast to the presence of both shear and bulk viscosity coefficients in the full DNS equation. A crucial difference between earlier analyses and our present discussion is that Damour's \cite{Damour:1979wya} or Padmanabhan's \cite{Padmanabhan:2010rp} calculations are performed on the null surface $r = r_h$, while our analysis is performed on a timelike hypersurface $r = r_c (>r_h)$. 
Another key difference is the nature of the kinematic viscosity. As discussed in length by Padmanabhan \cite{Padmanabhan:2010rp}, the viscous tensor in context to null surface dynamics is defined by a combination of Christoffel symbols which vanish in a local inertial frame. In our analysis since the base metric is Rindler in form, the viscosity is an observer dependent quantity and has the form $\eta = r_c$. The relevant observer one can think in this case is an accelerated one. This idea will be illuminated in the next section.


Let us now discuss the crucial aspects of the difference between the NS and DNS equations. We note that they are all invariant under the same scaling laws, represented by (\ref{set1}) (see Appendix \ref{App1} for a detailed derivation of the scale invariance). It might be useful to point out that, in principle, there may be higher order corrections to both the equations (typical correction terms are introduced in Section $3$ of \cite{Bredberg:2011jq}). But in all cases, one can check that these corrections are scaled away in the hydrodynamic limit. Therefore both the NS and DNS are equally allowed equations under these mentioned scalings. Although they are preserved under the same scaling laws, there is a crucial difference between them. NS equation is originally derived in flat space and so the reasonable derivative operator is the convective one, which is reflecting in the first two terms in (\ref{1.09}). Whereas, the DNS equation was originally obtained by the projection of Einstein field equations onto a null surface and hence is related to curved space-time. In this situation, the proper derivative operation (which deals with the variations of the different fluid variables) is taken care by the Lie derivative. Therefore the convective derivative is replaced in the fluid equation by the Lie derivative, thereby leading to the DNS equation. This is reflected in the first three terms of (\ref{1.10}). One can check that the Lie derivative of $v_i$ along $\xi^{\mu} = (1,v^j)$ is given by these three terms; i.e. $\pounds_{\xi}v_i=\xi^{\mu}\partial_\mu v_i + v_\mu\partial_i \xi^{\mu} = \partial_\tau v_i + v^j\partial_j v_i + (1/2)\partial_i v^2$. The last term is the main difference as it is not a part of the usual convective derivative.

Finally, we now mention that we have successfully achieved our goal to obtain the gravity dual of (restricted) DNS equation on a timelike hypersurface. The whole analysis is based on the idea of Arnowitt-Deser-Misner (ADM) formalism. In this formalism, one has to first choose the foliation and on this foliation, the Einstein field equations reduces to two groups of equations: one set contains the constraint equations and the other consists of dynamical equations of motion. The constraint equations constrain the initial data (i.e. values of induced metric $\gamma_{\mu\nu}$ and corresponding momentum in terms of $K_{\mu\nu}$) on the foliated surface, while the dynamical equations indicate how these variables change along the normal direction of the foliated surface. In the present case, the induced surface is the cutoff surface $\Sigma_c$, defined by $r=r_c$. We showed $\Sigma_c$ is described by the constraint Einstein equations (\ref{1.06}) which at the $\epsilon^2$ order leads to the incompressibility condition (\ref{1.07}) and at the $\epsilon^3$ order yields the fluid dynamical equation (\ref{1.08}) for the metric (\ref{1.01}). Therefore, the constraint Einstein equations are satisfied through $\mathcal{O}(\epsilon^3)$ as long as the incompressibility condition along with the fluid equation are satisfied. Of course, how the cutoff surface evolves will be determined by the dynamical Einstein equations of motion. Note that here the evolution will be along the radial direction. Hence the metric (\ref{1.01}) can be considered as a solution of Einstein's equations in presence of a particular type of bulk matter through $\epsilon^3$ order provided Eqs. (\ref{1.07}) and (\ref{1.08}) are satisfied. This indicates that the fluid equations, which are defined on the cutoff surface, are consistent with the constraint Einstein equations. The same argument is also considered in the original work \cite{Bredberg:2011jq}.  Of course, in that case the metric was a vacuum solution of the Einstein equations of motion.

\section{\label{pert}Perturbative approach}
In this section, we shall study the same topic in a slightly different approach which will again vivid the observer dependence more prominently.
Here the required geometry will be constructed in a perturbative manner by including the terms as an order by order correction to the leading order Rindler metric. Here the perturbative expansion will be taken in the hydrodynamic scaling parameter $\epsilon$ serving as corrections to the leading order flat metric. The whole process will be done under a particular choice of gauge in the metric coefficients. In this situation the relevant stress tensor and the extrinsic curvature will have corrections which we shall consider up to the linear order.

We consider a small perturbation of the metric coefficients as $\delta g_{A B} = h_{A B}$ under the following gauge:
\begin{align}
h_{rr}\,=\,h_{r\tau}\,=\,h_{ri}\,=\,0~.
\label{1.11}
\end{align}
The above is valid on any $r=$ constant surface, not necessarily on $r=r_c$. Along these we impose a boundary condition
\begin{align}
h_{A B}(r_c) = 0~.
\label{1.12}
\end{align}
So this is an asymptotic condition which says that the perturbations vanish on the cut-off surface. It must be mentioned that a similar gauge choice and boundary condition were also adopted in \cite{Bredberg:2010ky} for an identical context.
Let us now concentrate on the corrections to the stress tensor. The leading correction to the extrinsic curvature on the cutoff timelike hypersurface $\Sigma_c$ is given by,
\begin{align}
\delta K_{\mu \nu} =  K^{1}_{\mu \nu} 
= \sqrt{h} \partial_r h_{\mu \nu}~.
\label{1.13}
\end{align}
Thus, the leading correction to the Brown-York stress tensor is derived to be \cite{Bredberg:2010ky}
\begin{align}
\delta T^{\textrm{(BY)}}_{\mu \nu}(x^\mu, r_c) = T^{\textrm{(BY)} \, 1}_{\mu \nu}(x^\mu, r_c) = \sqrt{h}(-\partial_r h_{\mu \nu} + \gamma_{\mu \nu} \gamma^{\alpha \beta} \partial_r h_{\alpha \beta})~.
\label{1.14}
\end{align}
Therefore, now our relevant Brown-York stress tensor is the leading order Eq. (\ref{1.02}) plus the above first order correction to it. Eq. (\ref{1.02}) is to be evaluated from the base metric (Rindler form) while the above correction corresponds to perturbations onto the base metric.

With the gauge choice made in (\ref{1.11}) and including the tensor modes sourced by an appropriately defined bulk energy-momentum tensor $T_{A B}$ in the same spirit as in Section (\ref{dns}), the metric modifies to: 
\begin{align}
ds_{p+2}^2 &= g_{A B}^{(0)} dx^A dx^B + h_{A B} dx^A dx^B
\nonumber
\\
&= -rd\tau^2 + 2d\tau dr+dx_idx^i \nonumber \\
&+\bigg(1-\frac{r}{r_c}\bigg)v_i dx^i d\tau \nonumber \\
&+\bigg(1-\frac{r}{r_c}\bigg)\bigg[(v^2+P)d\tau^2 + \frac{v_i v_j}{r_c}dx^i dx^j + \tilde{a_1} \frac{v^2 \delta_{ij}}{r_c}dx^i dx^j + \tilde{a_2} (\partial_i v_j + \partial_j v_i) dx^i dx^j \bigg] \nonumber \\
&-\frac{r^2-r_c^2}{r_c} \partial^2 v_i dx^i d\tau +\dots
\label{1.15}
\end{align}
with the first line in the metric (\ref{1.15}) serving as the base or zeroth order metric $g_{A B}^{(0)}$ and the consequent lines serving as perturbations $h_{A B}$. 
For this modified metric (\ref{1.15}), the non zero contributions to the Ricci tensor $R_{A B}$ appear at $\mathcal{O}(\epsilon^2)$ and is derived using the explicit relations in Appendix (\ref{App2}) to be:
\begin{align}
R^{(2)}_{r \tau} &= - \frac{1}{r} R^{(2)}_{\tau \tau} = \frac{\tilde{a_1}}{4 r_c^2} p v^2 \,, \label{2.02a} \\
R^{(2)}_{i j} &= \frac{\tilde{a_1}}{2 r_c^2} v^2 \delta_{i j} + \frac{(\tilde{a_2} - 1)}{2 r_c} (\partial_i v_j + \partial_j v_i) \,, \label{2.02b}
\end{align}
where $\delta_{i i} = p$ refers to the number of angular dimensions. Correspondingly, the traceless bulk matter tensor $T_{A B}$  has non zero components up to $\mathcal{O}(\epsilon^2)$ of the following form:
\begin{align}
\kappa T^{(2)}_{\tau r} = R^{(2)}_{r \tau} &=  \frac{\tilde{a_1}}{4 r_c^2} p v^2 \,, \label{2.021} \\
\kappa T^{(2)}_{\tau \tau} = R^{(2)}_{\tau \tau} &= - \frac{\tilde{a_1}}{4 r_c^2} p r v^2 \,, \label{2.022} \\
\kappa T^{(2)}_{i j} = R^{(2)}_{i j} &= \frac{\tilde{a_1}}{2 r_c^2} v^2 \delta_{i j} + \frac{(\tilde{a_2} - 1)}{2 r_c} (\partial_i v_j + \partial_j v_i) \,. \label{2.023}
\end{align}

Now, on the timelike hypersurface $r=r_c$, the induced metric is flat:
\begin{align}
\gamma_{\mu \nu}dx^{\mu} dx^{\nu} = -r_cd\tau^2 + dx_i dx^i~.
\label{1.16}
\end{align}
For the zeroth order metric $g_{A B}^{(0)}$, the Brown-York stress tensor is given by (\ref{1.02}), while for the perturbed metric $h_{A B}$, the corrections to the stress tensor is calculated using (\ref{1.14}), thus giving:
\begin{align}
T_{\mu \nu}^{\textrm{(BY)}}dx^{\mu}dx^{\nu} &= \,\,T_{\mu \nu}^{\textrm{(BY)} \, 0}dx^{\mu}dx^{\nu} + T_{\mu \nu}^{\textrm{(BY)} \, 1}dx^{\mu}dx^{\nu} 
\nonumber
\\
&= \frac{dx^2_i}{2 \sqrt{r_c}} + (\tilde{a_1} p +1) \frac{v^2}{\sqrt{r_c}} d\tau^2 - 2\frac{v_i}{\sqrt{r_c}} dx^i d\tau \nonumber \\
&+ \frac{(v_iv_j + P\delta_{ij} + \tilde{a_1} (1-p) v^2 \delta_{ij})}{r^{3/2}_c} dx^i dx^j+ \tilde{a_2} \frac{(\partial_i v_j + \partial_j v_i)}{\sqrt{r_c}} dx^i dx^j + \mathcal{O}(\epsilon^3)~.
\label{1.17}
\end{align}
The constraint equations (\ref{1.06}) when satisfied on the cutoff hypersurface $\Sigma_c$, leads to the first nontrivial equation at order $\epsilon^2$ and with the free index as $\tau$, the incompressibility condition (\ref{1.07}).
The same constraint equations with the free index now assuming the angular variables $x^i$, we obtain at order $\epsilon^3$: 
\begin{eqnarray}
&&\partial_{\tau} v_i + v^j \partial_j v_i + \tilde{a_1} (1-p) \partial_i v^2 + \partial_i P + \tilde{a_2} \eta \partial^2 v_i 
\nonumber
\\
&&= \frac{\tilde{a_1}}{2 r_c} v_i v^2 + \frac{(\tilde{a_2} - 1)}{2} (v^j \partial_i v_j + v^j \partial_j v_i) \,,
\label{1.18}
\end{eqnarray}
where the kinematic viscosity is given by the formula $\eta = r_c$.

On assumption of the values $\tilde{a_1} = 0$ and $\tilde{a_2} = -1$, i.e. considering contributions solely from the tensor mode $(\partial_i v_j + \partial_j v_i) dx^i dx^j$, (\ref{1.18}) reduces to:
\begin{align}
\partial_{\tau} v_i + 2 v^j \partial_j v_i + \frac{1}{2} \partial_i v^2 + \partial_i P - \eta \partial^2 v_i = 0~.
\label{1.18a}
\end{align}
With the redefinition of the time co-ordinate introduced in Section (\ref{dns}), i.e., $\partial_{\tau} + v^j \partial_j \to \partial_{\tau}$, (\ref{1.18a}) takes up the form:
\begin{align}
\partial_{\tau} v_i +  v^j \partial_j v_i + \frac{1}{2} \partial_i v^2 + \partial_i P - \eta \partial^2 v_i = 0~,
\label{1.18b}
\end{align}
which is again the (restricted) DNS equation.

As in Section (\ref{dns}), we shall show that the bulk matter tensor $T_{A B}$ defined through relations (\ref{2.021})-(\ref{2.023}) with the above choices, i.e. $\tilde{a_1} = 0$ and $\tilde{a_2} = -1$, leads to the incompressibility condition independently of the constraint equation at $\mathcal{O}(\epsilon^2)$. Keeping in mind the traceless condition (\ref{1.02c}) imposed on $T_{A B}$, we obtain using (\ref{2.023}):
\begin{align}
T = g^{A B \, (0)} T_{A B}^{(2)} = - \frac{1}{r_c \kappa} \delta^{i j} (\partial_i v_j + \partial_j v_i) = 0 \implies \partial_i v^i = 0 \,.
\end{align}
Here also, like earlier, the Hamiltonian constraint equation (\ref{JHEP2}) is again automatically satisfied as the right hand side is of $\mathcal{O}(\epsilon^4)$.


It may be noted that the present analysis and results are based on the choice of the fixed gauge conditions (\ref{1.11}). Therefore it seems that the analysis is in contradiction to the general belief that physical results and quantities {\it must be} gauge invariant. But we shall argue that this is not at all unphysical in the present case; moreover such a feature, in the context of gravity, is not new at all. We remember that in gravity, most of the physical quantities, like temperature of the horizon, entropy of a black hole, etc. are observer dependent. Even the notion of event horizon is indeed observer dependent -- a freely falling frame will not see the horizon and consequently with respect to this frame the concept of temperature and entropy do not arise; whereas the outside static observer will see all of them. It has a close relation with the Unruh effect \cite{Unruh:1976db} which is perceived by a particular frame, known as Rindler frame. The same is reflected in the calculation of entropy by the Cardy formula where the diffeomorphism vectors are chosen by imposing certain relevant asymptotic gauge choice on the metric coefficients. These choices are usually done by imposing the asymptotic symmetries on the spacetime metric. It implies that certain degrees of freedom which were originally gauge degrees of freedom, now raise to true degrees of freedom and lead to black hole entropy. This is due to the particular choice of the gauge conditions which preserves certain symmetries of the metric viewed by a particular observer (these points are greatly discussed in \cite{Majhi:2012tf, Majhi:2013jpk}). 
Now in the present case the gauge conditions (\ref{1.11}) are chosen in such a way that the metric remains invariant (i.e. retains the Rindler form) on the cut-off surface $r=r_c$. Therefore it is evident that there exists an observer (Rindler, in this case), which confirms the invariance of its metric structure by choosing these gauge conditions, with respect to which these degrees of freedom due to perturbation do not contribute to the induced surface; whereas the rest do contribute and one finds the gravity dual of DNS fluid on this particular timelike induced surface. Since we are finding the gravity dual of a fluid, such an observer dependent feature may be unavoidable. 


\section{\label{near}Connection to near horizon limit approach}

The solutions (\ref{1.01}) and (\ref{1.15}) in the hydrodynamic $\epsilon$ expansion can be shown to be mathematically equivalent to a corresponding nonlinear version of a near-horizon expansion, similar in spirit to Section $8$ of \cite{Bredberg:2011jq}. For that we take the boundary on the accelerating surface to be $r = \tilde{r}_{c}$
such that $r \leq \tilde{r}_{c}$. The near-horizon limit is $\tilde{r}_{c} \to 0$. We make the transformations:
\begin{equation}
r = \tilde{r}_{c} \hat{r}~; \,\,\,\ \tau = \frac{\hat{\tau}}{\tilde{r}_{c}}~.
\label{New1}
\end{equation}

Let us first concentrate on the metric (\ref{1.01}). Under the transformations (\ref{New1}), it takes a very interesting form:
\begin{eqnarray}
ds_{p+2}^2 &&= -\frac{\hat {r}}{\tilde{r}_{c}} d\hat{\tau}^2 + \Big[2 d\hat{\tau} d\hat{r} + dx_i dx^i - 2(1-\hat{r})v_i dx^i d\hat{\tau} + (1-\hat{r}) (v^2+2P)d\hat{\tau}^2\Big]\nonumber \\
&& + \tilde{r}_{c}\Big[(1-\hat{r}) v_i v_j dx^i dx^j - 2v_i dx^i d\hat{r} + (v^2+2P)d\hat{\tau}d\hat{r} + (1-\hat{r}^2) \partial^2v_i dx^i d\hat{\tau} \nonumber \\
&& - (1-\hat{r}) \big(\tilde{a_1} v^2 \delta_{ij} dx^i dx^j + \tilde{a_2} (\partial_i v_j + \partial_j v_i) dx^i dx^j \big)\Big] + \mathcal{O}(\tilde{r}_{c}^2)~. 
\end{eqnarray}
Note that the above form can be interpreted as an expansion in the power series of $\tilde{r_c}$ which solves the Einstein equations (\ref{1.022a}) to $\mathcal{O}(\tilde{r}_{c})$. In addition it satisfies the relations (\ref{1.07}) and (\ref{1.10}), for appropriately defined values of $\tilde{a_1}$ and $\tilde{a_2}$ as discussed. Therefore one can conclude that the hydrodynamic expansion is mathematically equivalent to the near horizon expansion approach.

Similar conclusion can also be drawn for the metric (\ref{1.15}) discussed in Section (\ref{pert}). Under the transformations (\ref{New1}) it also takes an equivalent form like the near horizon expansion. The explicit expression is
\begin{eqnarray}
ds_{p+2}^2 &&=-\frac{\hat {r}}{\tilde{r}_{c}} d\hat{\tau}^2 + \Big[2 d\hat{\tau} d\hat{r} + dx_i dx^i - 2(1-\hat{r})v_i dx^i d\hat{\tau} + (1-\hat{r}) (v^2+2P)d\hat{\tau}^2\Big]  \nonumber \\
&& + \tilde{r}_{c}\Big[(1-\hat{r}) v_i v_j dx^i dx^j + (1-\hat{r}^2) \partial^2v_i dx^i d\hat{\tau} \nonumber \\
&&-(1-\hat{r}) \big(\tilde{a_1} v^2 \delta_{ij} dx^i dx^j + \tilde{a_2} (\partial_i v_j + \partial_j v_i) dx^i dx^j \big)\Big] + \mathcal{O}(\tilde{r}_{c}^2)~. 
\end{eqnarray}
Thus again the near-horizon expansion that solves the Einstein equations to $\mathcal{O}(\tilde{r}_{c})$ is equivalent to the hydrodynamic expansion approach.
 It has been checked that we are led to the same equations (\ref{1.07}) and (\ref{1.18b}), thus establishing the mathematical equivalence of the perturbative expansion in the hydrodynamic scaling parameter $\epsilon$ and the near-horizon expansion in $\tilde{r}_{c}$. 
\section{\label{disc}Discussions}
1.  {\it{Comments on the mode $\partial{_{_(i}}v{_{j_)}} dx^i dx^j$}}:
In both the frameworks, the tensor mode $(\partial_i v_j + \partial_j v_i) dx^i dx^j$ is crucially seen to generate a $\partial_i v^2$ term and this hints at a mathematical structure very similar to the DNS equation. As discussed in the penultimate paragraph of Section \ref{dns}, the (restricted) DNS equation (\ref{1.10}) contains the \textit{Lie derivative} of $v_i$ along $\xi^{A} = (1, 0, v^j)$ (i.e. the first three terms in (\ref{1.10})) while the NS equation contains the standard \textit{convective derivative}. This feature prompts a natural question as to whether the metric coefficient $(\partial_i v_j + \partial_j v_i) dx^i dx^j$, an $\mathcal{O}(\epsilon^2)$ term, could be generated from a corresponding \textit{Lie variation} of the base Rindler metric with respect to $\xi^{A} = (1, 0, v^j)$. To this end, using the relation:
\begin{align}
\pounds_{\xi} g_{i j} = \xi^A \partial_A g_{i j} + g_{j A} \partial_i \xi^A + g_{i A} \partial_j \xi^A ~,
\label{c1}
\end{align}
we obtain at $\mathcal{O}(\epsilon^2)$,
\begin{align}
\pounds_{\xi} g_{i j} \bigg|_{\mathcal{O}(\epsilon^2)} = g_{j k}^{(0)} \partial_i \xi^k + g_{i k}^{(0)} \partial_j \xi^k = \partial_i v_j + \partial_j v_i \,,
\label{c2}
\end{align}
which is clearly seen to be the case. In the above, $\pounds_{\xi}$ is the Lie derivative along $\xi^a$. It clearly justifies the inclusion of this tensor mode in the metric to obtain the DNS, and vice versa, in the present analysis. The same comment is also valid for the perturbative approach.

\vskip 2mm
\noindent
2. {\it Covariant conservation of the bulk matter tensor $T_{A B}$}:
The traceless energy momentum tensor $T_{A B}$ that eventually sources the mode $(\partial_i v_j + \partial_j v_i) dx^i dx^j$ as seen from relation (\ref{1.08c}) is covariantly conserved in the bulk to $\mathcal{O}(\epsilon^2)$, i.e. to the order the bulk solution is presented:
\begin{align}
\nabla_A T^{A B} = 0 \,.
\label{c3}
\end{align}
Since the only non-zero component is $T_{ij}$, an explicit calculation reveals that $\nabla_AT^{Ai}=\mathcal{O}(\epsilon^3)$ while others vanish identically for $\mathcal{O}(\epsilon^2)$ metric coefficients. 
Thus, we also show that our energy-momentum tensor $T_{A B}$ (\ref{1.08c}) has a zero covariant divergence and the corresponding mode it sources, leads to a generalized metric that solves the matter equations of motion (\ref{1.02d}). This is also true for the perturbative approach adopted here.

\vskip 2mm
\noindent
3. {\it Fate of the energy conditions}: Given an appropriately defined, covariantly conserved energy momentum tensor $T_{A B}$ characterizing the bulk matter, the strong energy condition (SEC) is the satisfaction of the inequality:
\begin{align}
T_{A B} v^A v^B \geq \frac{1}{2} T^A_A v^B v_B~,
\label{C1}
\end{align}
for all timelike $v^A$. A corresponding statement called the weak energy condition (WEC) is the inequality:
\begin{align}
T_{A B} v^A v^B \geq 0~.
\label{C2}
\end{align}
For a traceless energy momentum tensor (\ref{1.02c}), the SEC (\ref{C1}) essentially reduces to the WEC (\ref{C2}). Now for our case (\ref{1.08c}), the inequality computes to be:
\begin{align}
T_{A B} v^A v^B = T_{i j}^{(2)} v^i v^j = \frac{\tilde{a_2}}{2 \kappa r_c} (\partial_i v_j + \partial_j v_i) v^i v^j = - \frac{1}{2 \kappa r_c} v^i \partial_i v^2 \sim \mathcal{O}(\epsilon^4)\,. \quad (\because \tilde{a_2} = -1)
\label{C3}
\end{align}
On grounds of the argument that the bulk solution is presented to $\mathcal{O}(\epsilon^2)$ and the fact that the quantity (\ref{C3}) is of $\mathcal{O}(\epsilon^4)$ (and which consequently allows us to neglect it) automatically saturates the inequality (\ref{C2}), thereby trivially satisfying the energy condition(s). 

For the dominant energy condition (DEC), in addition to the WEC, the stress-tensor has to satisfy 
\begin{equation}
T_{AB}T^{B}_{C}v^Av^C\leq 0~.
\label{BC4}
\end{equation}
In the present case, since the only non-vanishing components of $T_{AB}$ is $T_{ij}$ and is of $\epsilon^2$ order, the above is automatically satisfied by the earlier argument.

\vskip 2mm
\noindent
4. {\it Possible form of matter action}: It would be interesting to find an action which leads to our matter energy momentum tensor (\ref{1.021})--(\ref{1.023}). Remember that in our present situation the only non-vanishing components of the energy momentum tensor are $T_{ij}^{(2)}$ with $\tilde{a}_1=0$ and $\tilde{a}_2=-1$, which consequently leads to the DNS equation (\ref{1.10}). In general, given a matter stress-tensor, it is not always easy to find the corresponding appropriate action. Here we shall attempt to construct a possible matter action for the bulk stress tensor at hand. 
We note that the metric component with coefficient $\tilde{a}_2$ in (\ref{1.01}) is liable for the non-vanishing component $T_{ij}^{(2)}$. We have already shown that this component is generated by the Lie variation of $g_{ij}$ along the direction $\xi^A$ (which is also of the order $\epsilon^2$). Therefore, roughly the energy-momentum tensor could be obtained from the following variation of the matter action:
\begin{equation}
\delta A_m=\frac{1}{4\kappa r_c}\int d^{p+2}x\,\delta g^{ij}(\partial_iv_j+\partial_jv_i)~.
\label{R1}
\end{equation}
Hence the action can be of the form:
\begin{equation}
A_m = \frac{1}{4\kappa r_c}\int d^{p+2}x\,g^{ij}(\partial_iv_j+\partial_jv_i)~.
\label{R2}
\end{equation}
To make it covariant, we propose the following form of the matter action:
\begin{equation}
A_m^{\textrm{(prop)}} = \frac{1}{4\kappa r_c}\int d^{p+2}x\,\sqrt{-g}\,\sigma^{AB}(\nabla_Av_B+\nabla_Bv_A)~,
\label{R3}
\end{equation}
where $\sigma_{AB}$ is the transverse metric on the ($x^i-x^j$) space and $v^A= (1,0,v^i)$. Since we are interested upto $\mathcal{O}(\epsilon^2)$ in the energy-momentum tensor, one can check that the above reduces to (\ref{R2}) at the required order. Since $\sigma_{AB}$ is the transverse metric, the only non-vanishing components are $\sigma_{ij}$. Therefore, we have
\begin{equation}
\sigma^{AB}(\nabla_Av_B+\nabla_Av_B) = 2\sigma^{ij}\nabla_iv_j = 2\sigma^{ij}(\partial_iv_j - \Gamma^\tau_{ij} - \Gamma^k_{ij}v_k)
\label{R4}
\end{equation}
which for our metric (\ref{1.01}) reduces to:
\begin{equation}
\sigma^{AB}(\nabla_Av_B+\nabla_Av_B) = 2g^{ij}\partial_iv_j + \mathcal{O}(\epsilon^3)~.
\label{R5}
\end{equation}
On the other hand for the action to be of the order $\epsilon^2$, we have $\sqrt{-g}=1$. Hence at order $\epsilon^2$, (\ref{R2}) and (\ref{R3}) are equivalent.

\vskip 2mm
\noindent
5. {\it Consistency with second law of thermodynamics for DNS}: To establish the fact that the (restricted) DNS equation (\ref{1.10}) is in fact consistent with the second law of thermodynamics, we have to show that the entropy production in such a fluid is positive definite. To this end, we consider the mass conservation and energy conservation equations along with the first law of thermodynamics for this DNS fluid. In our derivation, we work with energy and mass densities that are eventually integrated over the entire volume $\mathcal{V}$ to obtain the total energy and mass, respectively. Surface integrals, containing the velocity parameter $v_i$, over the closed boundary surface $\mathcal{S}$ of this volume vanish as the velocity field $v_i$ is identically zero at the boundary. Also we shall employ the incompressibility condition (\ref{1.07}) (which is essentially the mass conservation law) as and when required.

Keeping in mind the above arguments which we shall use in our calculations, we first turn our attention to the energy conservation for the dissipative (restricted) DNS fluid. The total energy $E_{\textrm{tot}}$ of the fluid under study is a sum of its kinetic energy $E_{\textrm{kin}}$, the internal energy $E_{\textrm{int}}$ and the energy due to heat flux transfer $E_{\textrm{h.f.}}$, i.e.:
\begin{equation}
E_{\textrm{tot}} = E_{\textrm{kin}} + E_{\textrm{int}} + E_{\textrm{h.f.}}~,
\label{R6}
\end{equation}
with the corresponding conservation law which reads:
\begin{equation}
\frac{D E_{\textrm{tot}}}{D \tau} = 0 \,,
\label{R7}
\end{equation}
where $\frac{D}{D \tau} = \frac{\partial}{\partial \tau} + v^j \partial_j = \partial_{\tau} + v^j \partial_j$ is the usual convective derivative. For calculating the kinetic energy contribution, we define the kinetic energy density $\epsilon_{\textrm{kin}} \equiv \frac{1}{2} v^2$. The rate of change of $\epsilon_{\textrm{kin}}$ is:
\begin{equation}
\frac{\partial \epsilon_{\textrm{kin}}}{\partial \tau} = \frac{\partial}{\partial \tau} \bigg( \frac{1}{2} v^2 \bigg) = v^i \frac{\partial v_i}{\partial \tau}~.
\label{R8}
\end{equation}
Recalling our (restricted) DNS fluid dynamical equation (\ref{1.10}), we have:
\begin{equation}
\partial_{\tau} v_i = - v^j \partial_j v_i - \frac{1}{2} \partial_i v^2 - \partial_i P + \eta_{\textrm{eff}} \partial^2 v_i~.
\label{R9}
\end{equation}
On using the fluid equation, (\ref{R8}) is now expressed as: 
\begin{equation}
\frac{\partial \epsilon_{\textrm{kin}}}{\partial \tau} = - v^i \partial_i v^2 - v^i \partial_i P + \eta_{\textrm{eff}}\, v^i \partial^2 v_i~.  
\label{R10}
\end{equation}
The viscous stress tensor is defined is defined as $\sigma_{i k}^{\prime} \equiv \eta_{\textrm{eff}}(\partial_k v_i + \partial_i v_k)$. Thus, we have the relation:
\begin{equation}
\partial^k \sigma^{\prime}_{i k} = \eta_{\textrm{eff}} \partial^2 v_i~.
\label{R11}
\end{equation}
Then on using (\ref{R11}), (\ref{R10}) simplifies to the expression:
\begin{equation}
\partial_{\tau} \epsilon_{\textrm{kin}} = -\partial_i \bigg[ 2 \epsilon v^i + P v^i + v^k \sigma^{\prime}_{i k} \bigg] - \sigma^{\prime}_{i k} \partial_k v_i~.
\label{R12}
\end{equation}
Integrating (\ref{R12}) over the volume $\mathcal{V}$, we obtain:
\begin{align}
\partial_{\tau} \int_{\mathcal{V}} dV \epsilon_{\textrm{kin}}  &= - \int_{\mathcal{V}} dV \partial_i \bigg[ 2 \epsilon v^i + P v^i + v^k \sigma^{\prime}_{i k} \bigg] - \int_{\mathcal{V}} dV \sigma^{\prime}_{i k} \partial_k v_i \nonumber \\
&= -\oint_{\mathcal{S}} dS_i \big[ 2 \epsilon v^i + P v^i + v^k \sigma^{\prime}_{i k} \bigg] - \int_{\mathcal{V}} dV \sigma^{\prime}_{i k} \partial_k v_i ~.
\label{R13}
\end{align}
Now, since the velocity field $v_i$ vanishes on the boundary $\mathcal{S}$, the first (surface) integral vanishes identically. Thus, we have,
\begin{align}
\partial_{\tau} \int_{\mathcal{V}} dV \epsilon_{\textrm{kin}} &= -\int_{\mathcal{V}} dV \sigma^{\prime}_{i k} \partial_k v_i = -\frac{1}{2} \int_{\mathcal{V}} dV \sigma^{\prime}_{i k} (\partial_k v_i + \partial_i v_k) \nonumber \\
&= -\frac{1}{2} \int_{\mathcal{V}} dV \sigma^{\prime}_{i k} \frac{\sigma^{\prime}_{i k}}{\eta_{\textrm{eff}}} = -\frac{1}{2 \eta_{\textrm{eff}}} \int_{\mathcal{V}} dV \sigma^2~.
\label{R14}
\end{align}

The mass conservation law (or, the equation of continuity) reduces to the incompressibility condition (\ref{1.07}), i.e., $\partial_i v^i = 0$. Now, from the first law of thermodynamics, we have:
\begin{align}
TdS = dU + PdV - \mu dN \nonumber \\
\implies dS = \frac{1}{T} dU - \frac{\mu}{T} dN ~.
\label{R15}
\end{align}
where the thermodynamic quantities have their usual meaning. On grounds of incompressibility, $dV = dN = 0$ and thus (\ref{R15}) reduces to (in terms of the entropy and internal energy densities):
\begin{align}
ds=\frac{1}{T} d\epsilon_{\textrm{int}}~. 
\label{R16}
\end{align}
In view of the equations (\ref{R6}), (\ref{R7}) and (\ref{R16}), we obtain the relation:
\begin{align}
\frac{D s}{D \tau} = -\frac{1}{T} \Big(\frac{D \epsilon_{\textrm{kin}}}{D \tau} + \frac{D \epsilon_{\textrm{h.f.}}}{D \tau}\Big)~.
\label{R17}
\end{align}
Now, we have $\frac{D \epsilon_{\textrm{kin}}}{D \tau} =  \partial_{\tau} \epsilon_{\mathrm{kin}}$ (the other part being a total derivative, vanishes on the surface $\mathcal{S}$ on integrating over the volume $\mathcal{V}$ as $v^i\partial_i\epsilon_{\textrm{kin}}=\partial_i(v^i\epsilon_{\textrm{kin}})$). A similar argument also holds for the second term of the above equation. Thus, in terms of the heat flux $q_j$, (\ref{R17}) gives:
\begin{align}
D_{\tau} s = -\frac{1}{T} \partial_{\tau} \epsilon_{\textrm{kin}} - \frac{1}{T} \partial_i q^i = \frac{1}{2 \eta_{\textrm{eff}}} \sigma^2 - \frac{1}{T} \partial_i q^i~,
\label{R18}
\end{align}
where we have used the relation (\ref{R14}) and $\partial_\tau\epsilon_{\textrm{h.f.}} = \partial_iq^i$. Using the relation $\frac{1}{T} \partial_i q^i = \partial_i \big( \frac{q^i}{T}\big) + \frac{q^i}{T^2}\partial_i T$, (\ref{R18}) modifies to:
\begin{align}
D_{\tau} s = \partial_{\tau} s + v^i \partial_i s = \frac{1}{2 \eta_{\textrm{eff}}} \sigma^2 - \partial_i \bigg( \frac{q^i}{T}\bigg) - \frac{q^i}{T^2}\partial_i T \nonumber \\
\implies \partial_{\tau} s + \partial_i \bigg( s v^i + \frac{q^i}{T} \bigg) = \frac{1}{2 \eta_{\textrm{eff}}} \sigma^2 - \frac{q^i}{T^2}\partial_i T~.
\label{R19}
\end{align}
Using the definition of the entropy current for dissipative fluids $s^i = s v^i + \frac{q^i}{T}$ and for the heat flux $q^i = - \kappa \partial^i T$, where $\kappa$ is the thermal conductivity ($\kappa \geq 0$), we obtain:
\begin{align}
\partial_{\tau} s + \partial_i s^i = \frac{1}{2 \eta_{\textrm{eff}}} \sigma^2 + \frac{\kappa}{T^2} (\partial^i T)(\partial_i T) = \frac{1}{2 \eta_{\textrm{eff}}} \sigma^2 + \frac{\kappa}{T^2} (\partial T)^2 > 0~,
\label{R20}
\end{align}
which shows that the entropy production in the (restricted) DNS fluid is positive and thus consistent with the second law of thermodynamics.

\vskip 2mm
\noindent
6. {\it Diffeomorphism symmetry of DNS}: It is worth pointing out that the NS equation possesses Galilean symmetry which is compatible with the idea of constructing the fluid equation at the Newtonian level. On the other hand, the DNS equation was originally obtained from a purely gravitational perspective. So far, it is known that the Einstein's equation when projected onto a null surface exhibits such a structure. In this paper, the same has been derived on a timelike hypersurface for the first time by projecting the Einstein's equation on it. So the origin of the DNS equation has an inherent curved spacetime concept. This is also supported by the fact that the DNS term is sourced by the Lie variation of the metric coefficient $g_{ij}$ and the convective derivative is seen to be replaced by the Lie derivative.  Therefore, whether the DNS equation is Galilean invariant, is a tricky question to ask. We know that in a curved background, a locally inertial frame can be constructed. So a better question to ask would be: is it Galilean invariant in a local inertial frame? The answer is ``yes''. This is because, in \cite{Padmanabhan:2010rp}, the author shows that the DNS term vanishes in a local inertial frame and one recovers the NS equation. Then automatically the reduced equation is Galilean invariant as it is the original NS equation. Therefore for the DNS equation, which is natural to a general frame, one needs to check its invariance under infinitesimal diffeomorphism $x^A\rightarrow x^A+\epsilon^A$ where $\epsilon^A$ is very small, say of the order $\epsilon$. One can observe that the Lie variation of the left hand side of Eq. (\ref{1.10}) creates terms which are higher than $\mathcal{O}(\epsilon^3)$. Now since we want to retain the scaling symmetry (\ref{set1}), these higher order terms are automatically scaled away in the hydrodynamic limit. Hence, the (restricted) DNS equation possesses diffeomorphism symmetry.

\vskip 2mm
\noindent
7. {\it Alternative interpretation of non-vanishing $R_{AB}$}:
 In our main analysis, we have argued that the metric (\ref{1.01}) can be seen as a solution to the Einstein's equations of motion in presence of a particular type of matter in the spacetime. We interpreted the non-vanishing component $R_{ij}^{(2)}$ as the matter stress-tensor. Below we try to explore another possible interpretation of this. 

Collectively it can be seen that the metric (\ref{1.01}) is a solution of the equations of motion 
\begin{equation}
R_{AB}-\frac{1}{2}Rg_{AB}+CP_{AB}=0~,
\label{7R1}
\end{equation} 
where $C$ is a constant and $P_{AB}$ is some tensor whose nonvanishing component is $P_{ij}^{(2)}$. This $P_{AB}$ is not due to the bulk matter (say), rather it is a part of some additional term in the gravitational action. It means here we are trying to find a ``modified'' theory of gravity whose action is given by the Einstein-Hilbert part plus some additional piece which leads to the required $P_{AB}$. Remember that the only non-vanishing component of this is $P_{ij}^{(2)}\sim \partial_iv_j+\partial_jv_i$. Let us now look at the possible choices of $P_{AB}$.

If we consider our gravitational theory to be a subclass of  Lanczos-Lovelock (LL) gravity, then $P_{AB}$ will contain terms which is a product of either two of Ricci scalar, Ricci tensor or the Riemann tensor; i.e. $P_{AB}\sim R*R$. Now since $P_{AB}$ has to be $\mathcal{O}(\epsilon^2)$ and have the above structure (i.e. $P_{ij}^{(2)}\sim\partial_iv_j+\partial_jv_i$), the LL terms will not produce such terms (remember that for the metric (\ref{1.01}), all $R$, $R_{AB}$ vanish upto first order in $\epsilon$ while $R_{ABCD}\sim v_i$). Therefore the modification to Einstein's gravity can not be from a higher order theory. Next, we can try with the inclusion of the cosmological constant $\Lambda$ in the action, and then $P_{AB}\sim \Lambda g_{AB}$. In this case, to ensure a vanishing $P_{AB}^{(0)}$, we must choose $\Lambda=0$ and consequently $P_{AB}=0$ at all orders in $\epsilon$. Hence, the cosmological constant cannot serve our purpose. With all these failures, we now include a term $\sim \int d^{p+2}x\,\sqrt{-g}\,\Lambda(g_{ij})$ in the action. It will lead to a $P_{AB}$ of the following form:
\begin{equation}
P_{AB}\sim \Lambda(g_{ij})g_{AB}  + \frac{\partial\Lambda}{\partial g^{AB}}~.
\label{7R2}
\end{equation}
Now since $\Lambda$ is only a function of $g_{ij}$, the last term on the right hand side of the above equation will yield the non-vanishing components to be $\partial\Lambda/\partial g^{ij}$. In that case for our metric (\ref{1.01}), one must have $\Lambda g_{ij}+\partial\Lambda/\partial g^{ij}|_{\mathcal{O}(\epsilon^2)}\sim\partial_iv_j+\partial_jv_i$, while the others must vanish upto $\mathcal{O}(\epsilon^2)$. This is possible if $\Lambda$ is itself of $\mathcal{O}(\epsilon^2)$. One possible choice which fulfils the above criterion is $\Lambda(g_{ij}) = \partial_iv^i = g^{ij}\partial_iv_j$. Then the above $P_{AB}$ reduces to $P_{ij}=\partial\Lambda/\partial g^{ij}\sim\partial_iv_j+\partial_jv_i$ only, as $v_i$ satisfies the incompressibility condition. Hence the metric (\ref{1.01}) is a solution of Einstein's equations of motion with a varying ``cosmological parameter'' $\Lambda = \partial_iv^i$ which happens to be a function of the transverse metric $g_{ij}$. This idea is very close to the earlier interpretation as that of the bulk matter. The reason is that the cosmological constant is itself sometimes interpreted as some form of matter in cosmology.

In this regard, we want to bring to notice that a varying cosmological parameter is not a new concept. It has been followed in black hole thermodynamics as well, where $\Lambda$ is also a variable. In such a setting, $\Lambda$ is interpreted as the pressure term ($P$) which leads to the $VdP$ term in the first law of anti-De-Sitter (AdS) black hole, where $V$ is the volume of the black hole region. In this respect, it has been observed that the isotherms in the $P-V$ diagram are identical to the van der Waals gas system. Detailed analyses of such work can be found in \cite{Kastor:2009wy, Dolan:2011xt, Kubiznak:2012wp, Majhi:2016txt, Bhattacharya:2017hfj}.

The metric construction in the original work \cite{Bredberg:2011jq} is based on the boundary condition that the induced metric on the $r=r_c$ surface must be flat. One can check that our metric (\ref{1.01}) also satisfies the same boundary condition. Here we want to point out that this is not in contradiction with the above discussion where we gave an alternative interpretation that the metric is a solution of an Einstein-varying cosmological parameter theory. Recall that the cosmological parameter is of $\epsilon^2$ order and so it does not contribute to the zeroth order metric. Therefore, the induced metric at the zeroth order, which is the base metric of our theory, is naturally flat as the  perturbation hits at $\mathcal{O}(\epsilon^2)$ in the form of the cosmological parameter.

\section{\label{conc}Conclusions}
It was well known that the DNS equation governs the geometric data on any null surface in the spacetime. In our paper we have shown that a restricted version of the DNS equation arises even when the geometry under consideration is a timelike hypersurface. The restriction is in the form of the incompressibility condition and hence we call it an incompressible DNS equation. The geometry can be adapted to generate NS or DNS dynamics, with an appropriate bulk matter tensor needed to be defined for the latter case, as discussed.

In an alternative framework, we have shown that the geometry can also be considered as perturbations on a flat Rindler leading order metric. In both the frameworks, the tensor mode $(\partial_i v_j + \partial_j v_i) dx^i dx^j$ is crucially seen to generate a $\partial_i v^2$ term and thus hints at a mathematical structure very similar to the DNS equation. We have also shown that this tensor mode is generated due to an appropriately defined bulk matter in the spacetime and is characterized by an energy-momentum tensor with a certain constraint, as discussed. We also establish the mathematical equivalence of the hydrodynamic expansion method to that of the near horizon expansion. 


There are certain points that are needed to be looked into which may be of considerable interest in the present context. It has been observed that the metric of \cite{Bredberg:2011jq} (i.e. the metric (\ref{1.01}) with $\tilde{a}_1$ and $\tilde{a}_2$ set to zero) can also be obtained by giving a boost along $\tau$ and $x^i$ for constant $v^i$ and then imposing a linear shift in $r$ accompanied by a scaling of $\tau$ and applying these transformations on the Rindler spacetime. Finally on considering $v^i$ and $P$ as functions of $x^i$ and $\tau$ and expanding them around their background values in the hydrodynamic limit, leads to the required metric (for details, see \cite{Compere:2011dx}). It would be interesting if the new terms of this paper (i.e. terms with coefficients $\tilde{a}_1$ and $\tilde{a}_2$ of the metric (\ref{1.01})) can also be produced in a similar formalism. Moreover, how one can incorporate all higher order terms in the metric and its immediate consequences with respect to the DNS equation, must be addressed. For NS equation, the same has been discussed both in the non-relativistic \cite{Compere:2011dx} and relativistic \cite{Compere:2012mt} regimes. At the present moment, we leave this problem for future considerations and we hope to report these directions soon.


\acknowledgments
We thank the authors of \cite{Bredberg:2011jq} for clarifying several points about their work. We are grateful to Prahar Mitra for illuminating comments on an advanced version of the manuscript. The anonymous referee is also greatly acknowledged for making several critical, but constructive comments.


\vskip 5mm
\section*{Appendices}
\appendix
\section{{\label{App1}}Scale invariance of restricted DNS Equation (\ref{1.10})}
The proof for the scale invariance of the incompressible NS equation is well established mathematically. Here we show that the restricted DNS equation (\ref{1.10}) is also scale invariant under the same scaling laws (\ref{set1}) and (\ref{set2}) as that of the incompressible NS equation. 

To establish the proof of scale invariance of (\ref{1.10}), we first consider the space domain to scale as $\tilde{\Lambda}=\lambda\Lambda$, with $\lambda \in \mathbb{R}^+$. $\Lambda$, in the present case, stands for $x^i$. Correspondingly, we assume that the time and pressure scale as $\tilde{\tau}=\lambda^{1-h}\tau$ and $\tilde{P}(\tilde{x}^i,\tilde{\tau})=\lambda^m P(\tau,x^i)$, respectively, with the values of $h$ and $m$ to be determined. Under these scalings, one finds
\begin{equation}
\tilde{v}_i(\tilde{x}^i,\tilde{\tau})=\lambda^h v_i(x^i,\tau)~.
\label{new2}
\end{equation}
We remember that $v^i$ is defined with respect to its arguments; i.e. $\tilde{v}_i = d\tilde{x}_i/d\tilde{\tau}$, and so on. We shall follow the same notation in our further analysis.

Thus we obtain the scaling relations of the terms in the restricted DNS equation (\ref{1.10}) to be as follows:
\begin{align}
\partial_{\tilde{\tau}} \tilde{v}_i(\tilde{x}^i,\tilde{\tau})&=\lambda^{2h-1} \partial_{\tau}v_i (x^i,\tau)~;
\nonumber
\\
\tilde{v}^j (\tilde{x}^i,\tilde{\tau}) \tilde{\partial}_j \tilde{v}_i(\tilde{x}^i,\tilde{\tau})&=\lambda^{2h-1} v^j(x^i,\tau)\partial_j v_i (x^i,\tau)~; 
\nonumber 
\\
\tilde{\partial}_i \tilde{v}^2(\tilde{x}^i,\tilde{\tau})&=\lambda^{2h-1} \partial_i v^2 (x^i,\tau)~; 
\nonumber 
\\
\tilde{\partial}_i \tilde{P}(\tilde{x}^i,\tilde{\tau})&=\lambda^{m-1}\partial_i P(x^i,\tau)~; 
\nonumber 
\\
\eta \tilde{\partial}^2 \tilde{v}_i(\tilde{x}^i,\tilde{\tau})&= \lambda^{h-2}\eta\partial^2 v_i(x^i,\tau)~,
\label{new3}
\end{align}
where $\eta$ is the kinematic viscosity and $i=1, \ldots p$. 
Now in order for Eq. (\ref{1.10}) to be scale invariant, each term of the equation should transform identically. Consequently, for $\eta \neq 0$, (\ref{new3}) tells that we must have $h-2 = 2h-1$ and $2h-1=m-1$. These relations imply $h = -1$ and $m=-2$. Therefore we have the following transformations:
\begin{align}
\tilde{x}^i&=\lambda x^i; \,\,\,\ \tilde{\tau}=\lambda^2\tau~;
\nonumber
\\
\tilde{v}_i(\tilde{x}^i,\tilde{\tau}) &= \lambda^{-1}v_i(x^i,\tau)~; 
\nonumber
\\
\tilde{P}(\tilde{x}^i,\tilde{\tau})&=\lambda^{-2}P(x^i,\tau)~;
\nonumber
\\
\partial_{\tilde{\tau}} \tilde{v}_i(\tilde{x}^i,\tilde{\tau})&=\lambda^{-3} \partial_{\tau}v_i (x^i,\tau)~;
\nonumber
\\
\tilde{v}^j (\tilde{x}^i,\tilde{\tau}) \tilde{\partial}_j \tilde{v}_i(\tilde{x}^i,\tilde{\tau})&=\lambda^{-3} v^j(x^i,\tau)\partial_j v_i (x^i,\tau)~; 
\nonumber 
\\
\tilde{\partial}_i \tilde{v}^2(\tilde{x}^i,\tilde{\tau})&=\lambda^{-3} \partial_i v^2 (x^i,\tau)~; 
\nonumber 
\\
\tilde{\partial}_i \tilde{P}(\tilde{x}^i,\tilde{\tau})&=\lambda^{-3}\partial_i P(x^i,\tau)~; 
\nonumber 
\\
\eta \tilde{\partial}^2 \tilde{v}_i(\tilde{x}^i,\tilde{\tau})&= \lambda^{-3}\eta\partial^2 v_i(x^i,\tau)~.
\label{new4}
\end{align}

Next we demand the DNS equation to remain invariant under some transformations of the fluid variables $v_i$ and $P$. For that we need to scale these fluid variables in such a way that each term in (\ref{1.10}) remains invariant. To find these relations, we first concentrate on the fourth equation of (\ref{new4}). Using the scaling of $\tau$, this can be re-expressed as
\begin{align}
 \partial_\tau \tilde{v}_i(\tilde{x}^i,\tilde{\tau}) = \partial_\tau\Big[\lambda^{-1}v_i(x^i,\tau)\Big]=\partial_\tau\Big[\lambda^{-1}v_i(\lambda^{-1}\tilde{x}^i,\lambda^{-2}\tilde{\tau})\Big]~,
\label{new5} 
\end{align}
where in the last step, the first equation of (\ref{new4}) has been used. This implies that the above term is invariant under the following scaling of the velocity field:
\begin{align}
\tilde{v}_i({\tilde{x}^i,\tilde{\tau}})= \lambda^{-1}v_i(\lambda^{-1}\tilde{x}^i,\lambda^{-2}\tilde{\tau})~.
\label{new6}
\end{align}
With this scaling, one can check that the other terms composed of $v_i$ only, also remain invariant. In a similar argument, it is easy to show that the pressure scales as
\begin{align}
\tilde{P}(\tilde{x}^i,\tilde{\tau})=\lambda^{-2}P(\lambda^{-1}\tilde{x}^i,\lambda^{-2}\tilde{\tau})~,
\label{new7}
\end{align}
which keeps the seventh equation of (\ref{new4}) invariant.

Thus we see that the pair of rescaled fluid variables $v_i^\lambda(x^i,\tau)$ and $P^\lambda(x^i,\tau)$, which are related to the old ones by the following relations 
\begin{align}
v_i^\lambda(x^i,\tau)&=\lambda^{-1}v_i(\lambda^{-1}x^i,\lambda^{-2}\tau)~;
\nonumber
\\
P^\lambda(x^i,\tau)&=\lambda^{-2}P(\lambda^{-1}x^i,\lambda^{-2}\tau)~, 
\end{align}
also solve the restricted DNS equation (\ref{1.10}) (in the above we have re-labelled $\tilde{x}^i$ and $\tilde{\tau}$ of (\ref{new6}) and (\ref{new7}) as $x^i$ and $\tau$, respectively). As stated earlier, the above fluid variables are defined with respect to its arguments, i.e., $v_i(\lambda^{-1}x^i,\lambda^{-2}\tau)=d(\lambda^{-1}x^i)/d(\lambda^{-2}\tau)$. 
Now with $\lambda^{-1}=\epsilon$, the amplitudes are scaled down by the parameter $\epsilon$ and leads to (\ref{set1}), thus recovering the same scaling laws (\ref{set1}) as that of the incompressible NS equation. As discussed in \cite{Bredberg:2011jq} for NS, typical corrections that can arise get scaled away in the hydrodynamic limit $\epsilon \to 0$ for (\ref{1.10}) as well.

\section{{\label{App2}}Explicit computation of the Ricci tensor $\mathbf{R}_\mathbf{\mu \nu}$ for (\ref{1.01})}
We explicitly present the computation of the Ricci tensor components $R_{\mu \nu}$ for a metric identical to (\ref{1.01}) but whose coefficients have now been pre-factored with suitable constants $a_i$. This clearly indicates as to how the various metric coefficients contribute to the Ricci tensor components at various orders in $\epsilon$ and the manner in which the pre-factors $a_i$ need to be chosen for solving the Einstein field equations, in vacuum or in presence of matter. To this end, we have the following metric:
\begin{align}
ds_{p+2}^2=&-rd\tau^2 + 2d\tau dr + dx_idx^i \nonumber \\
&-2 a_1 \bigg(1-\frac{r}{r_c}\bigg)v_i dx^i d\tau - 2 a_2 \frac{v_i}{r_c} dx^i dr \nonumber \\
&+\bigg(1-\frac{r}{r_c}\bigg)\bigg[ a_3 (v^2+2P)d\tau^2 + a_4 \frac{v_i v_j}{r_c}dx^i dx^j + a_5 \frac{v^2 \delta_{ij}}{r_c}dx^i dx^j \nonumber \\
&+ a_6 (\partial_i v_j + \partial_j v_i) dx^i dx^j\bigg] + a_7 \bigg(\frac{v^2}{r_c} + \frac{2P}{r_c}\bigg)d\tau dr + \mathcal{O}(\epsilon^3) \, . 
\label{A2.1} 
\end{align}

For the flat Rindler metric at $\mathcal{O}(\epsilon^0)$, the non-zero Christoffel symbols are:
\begin{align}
\Gamma^{r \, (0)}_{\tau r} = -\frac{1}{2} \, , \quad \Gamma^{r \, (0)}_{\tau \tau} = \frac{r}{2} \, , \quad \Gamma^{\tau \, (0)}_{\tau \tau} = \frac{1}{2} \, .
\label{A2.2}
\end{align}
From the expression of the Ricci tensor
\begin{align}
R_{\mu \nu} = \partial_{\alpha} \Gamma^{\alpha}_{\mu \nu} - \partial_{\nu} \Gamma^{\alpha}_{\mu \alpha} + \Gamma^{\alpha}_{\mu \nu} \Gamma^{\beta}_{\alpha \beta} - \Gamma^{\beta}_{\mu \alpha} \Gamma^{\alpha}_{\nu \beta}
\label{A2.3}
\end{align}
it is easy to see that all the components of the Ricci tensor vanish identically at the zeroth order in $\epsilon$, i.e.,
\begin{align}
R_{r r}^{(0)} = R_{r \tau}^{(0)} = R_{r i}^{(0)} = R_{\tau \tau}^{(0)} = R_{\tau i}^{(0)} = R_{i j}^{(0)} = 0 \, .
\end{align}

The non-zero Christoffel symbols in $\mathcal{O}(\epsilon)$ are listed as follows:
\begin{align}
\Gamma^{r \, (1)}_{r i} = \frac{a_1}{2 r_c} v_i \,, \quad \Gamma^{r \, (1)}_{\tau i} = -r \frac{a_1}{2 r_c} v_i \,, \quad \Gamma^{\tau \, (1)}_{\tau i} = -\frac{a_1}{2 r_c} v_i \,, \quad \Gamma^{i \, (1)}_{r \tau} = \frac{a_1}{2 r_c} v^i \,.
\label{A2.3a}
\end{align}
At the first order in $\epsilon$, the contributions to the Ricci tensor come from both the zeroth and first order Christoffel symbols. Keeping in mind the scaling of the partial derivatives as given in (\ref{set2}) and using (\ref{A2.3}) along with (\ref{A2.2}) and (\ref{A2.3a}), the various components of the Ricci tensor at $\mathcal{O}(\epsilon)$ are worked out explicitly to be:
\begin{align}
R_{r r}^{(1)} &= \partial_r \Gamma^{r \, (1)}_{r r} + \partial_i \Gamma^{i \, (0)}_{r r} - \partial_r \Gamma^{r \, (1)}_{r r} - \partial_r \Gamma^{\tau \, (1)}_{r \tau} - \partial_r \Gamma^{i \, (1)}_{r i} \nonumber \\
&+ \Gamma^{\alpha \, (0)}_{r r} \Gamma^{\beta \, (1)}_{\alpha \beta} + \Gamma^{\alpha \, (1)}_{r r} \Gamma^{\beta \, (0)}_{\alpha \beta} - \Gamma^{\beta \, (1)}_{r \alpha} \Gamma^{\alpha \, (0)}_{r \beta} - \Gamma^{\beta \, (0)}_{r \alpha} \Gamma^{\alpha \, (1)}_{r \beta} = 0 \, , \nonumber \\
R_{r \tau}^{(1)} &= \partial_r \Gamma^{r \, (1)}_{r \tau} + \partial_i \Gamma^{i \, (0)}_{r \tau} + \Gamma^{\alpha \, (0)}_{r \tau} \Gamma^{\beta \, (1)}_{\alpha \beta} + \Gamma^{\alpha \, (1)}_{r \tau} \Gamma^{\beta \, (0)}_{\alpha \beta} - \Gamma^{\beta \, (0)}_{r \alpha} \Gamma^{\alpha \, (1)}_{\tau \beta} - \Gamma^{\beta \, (1)}_{r \alpha} \Gamma^{\alpha \, (0)}_{\tau \beta} = 0 \, , \nonumber \\
R_{r i}^{(1)} &= \partial_r \Gamma^{r \, (1)}_{r i} + \partial_j \Gamma^{j \, (0)}_{r i} - \partial_i \Gamma^{\alpha \, (0)}_{r \alpha} + \Gamma^{\alpha \, (0)}_{r i} \Gamma^{\beta \, (1)}_{\alpha \beta} + \Gamma^{\alpha \, (1)}_{r i} \Gamma^{\beta \, (0)}_{\alpha \beta} - \Gamma^{\beta \, (0)}_{r \alpha} \Gamma^{\alpha \, (1)}_{i \beta} - \Gamma^{\beta \, (1)}_{r \alpha} \Gamma^{\alpha \, (0)}_{i \beta} = 0 \, , \nonumber \\
R_{\tau \tau}^{(1)} &= \partial_r \Gamma^{r \, (1)}_{\tau \tau} + \partial_i \Gamma^{i \, (0)}_{\tau \tau} + \Gamma^{\alpha \, (0)}_{\tau \tau} \Gamma^{\beta \, (1)}_{\alpha \beta} + \Gamma^{\alpha \, (1)}_{\tau \tau} \Gamma^{\beta \, (0)}_{\alpha \beta} - \Gamma^{\beta \, (0)}_{\tau \alpha} \Gamma^{\alpha \, (1)}_{\tau \beta} - \Gamma^{\beta \, (1)}_{\tau \alpha} \Gamma^{\alpha \, (0)}_{\tau \beta} = 0 \, , \nonumber \\
R_{\tau i}^{(1)} &= \partial_r \Gamma^{r \, (1)}_{\tau i} + \partial_j \Gamma^{j \, (0)}_{\tau i} - \partial_i \Gamma^{\alpha \, (0)}_{\tau \alpha} + \Gamma^{\alpha \, (0)}_{\tau i} \Gamma^{\beta \, (1)}_{\alpha \beta} + \Gamma^{\alpha \, (1)}_{\tau i} \Gamma^{\beta \, (0)}_{\alpha \beta} - \Gamma^{\beta \, (0)}_{\tau \alpha} \Gamma^{\alpha \, (1)}_{i \beta} - \Gamma^{\beta \, (1)}_{\tau \alpha} \Gamma^{\alpha \, (0)}_{i \beta} = 0 \, \nonumber \\
R_{i j}^{(1)} &= \partial_r \Gamma^{r \, (1)}_{i j} + \partial_l \Gamma^{l \, (0)}_{i j} - \partial_j \Gamma^{\alpha \, (0)}_{i \alpha} + \Gamma^{\alpha \, (0)}_{i j} \Gamma^{\beta \, (1)}_{\alpha \beta} + \Gamma^{\alpha \, (1)}_{i j} \Gamma^{\beta \, (0)}_{\alpha \beta} - \Gamma^{\beta \, (0)}_{i \alpha} \Gamma^{\alpha \, (1)}_{j \beta} - \Gamma^{\beta \, (1)}_{i \alpha} \Gamma^{\alpha \, (0)}_{j \beta} = 0 \,. \nonumber \\
\end{align}

The non-zero Christoffel symbols at $\mathcal{O}(\epsilon^2)$ are listed as follows:
\begin{align}
&\Gamma^{r \, (2)}_{i j} = (a_1 - a_2) \frac{r}{r_c} \partial_{(i}v_{j)} - a_1 \partial_{(i}v_{j)} + \frac{r}{2 r_c} \bigg{[}a_4 \frac{v_i v_j}{r_c}+ a_5 \frac{v^2 \delta_{ij}}{r_c}+ a_6 \partial_{(i}v_{j)}\bigg{]} \,, \nonumber \\
&\Gamma^{\tau \, (2)}_{r \tau} = - \frac{a_1 a_2}{2 r_c^2} v^2 \,, \quad \Gamma^{r \, (2)}_{r \tau} = - \frac{a_1 a_2}{2 r_c^2} r v^2 \,, \quad \Gamma^{r \, (2)}_{\tau \tau} = \frac{r}{4 r_c} a_3 (v^2 + 2P) \nonumber \\
&\Gamma^{i \, (2)}_{r i} = \frac{1}{2} \delta^{i j} \bigg{[} \bigg{(}-\frac{1}{r_c}\bigg{)}\bigg{(}a_4 \frac{v_i v_j}{r_c}+ a_5 \frac{v^2 \delta_{ij}}{r_c}+ a_6 \partial_{(i}v_{j)}\bigg{)} + \frac{a_2}{r_c} (\partial_j v_i - \partial_i v_j) \bigg{]} \nonumber \\
&\Gamma^{\tau \, (2)}_{\tau \tau} = \frac{1}{2 r_c} a_3 (v^2 + 2P) \,.
\label{A2.4}
\end{align}
Using (\ref{A2.2}), (\ref{A2.3a}) and (\ref{A2.4}), we obtain for the Ricci tensor components at $\mathcal{O}(\epsilon^2)$ to be:
\begin{align}
R_{r r}^{(2)} &= \partial_r \Gamma^{r \, (2)}_{r r} + \partial_i \Gamma^{i \, (1)}_{r r} + \partial_{\tau} \Gamma^{\tau \, (0)}_{r r} - \partial_r \Gamma^{r \, (2)}_{r r} - \partial_r \Gamma^{\tau \, (2)}_{r \tau} - \partial_r \Gamma^{i \, (2)}_{r i} \nonumber \\
&+ \Gamma^{\alpha \, (0)}_{r r} \Gamma^{\beta \, (2)}_{\alpha \beta} + \Gamma^{\alpha \, (1)}_{r r} \Gamma^{\beta \, (1)}_{\alpha \beta} + \Gamma^{\alpha \, (2)}_{r r} \Gamma^{\beta \, (0)}_{\alpha \beta} - \Gamma^{\beta \, (0)}_{r \alpha} \Gamma^{\alpha \, (2)}_{r \beta} - \Gamma^{\beta \, (1)}_{r \alpha} \Gamma^{\alpha \, (1)}_{r \beta} - \Gamma^{\beta \, (2)}_{r \alpha} \Gamma^{\alpha \, (0)}_{r \beta} = 0 \, , \nonumber \\
R_{r \tau}^{(2)} &= \partial_r \Gamma^{r \, (2)}_{r \tau} + \partial_i \Gamma^{i \, (1)}_{r \tau} + \partial_{\tau} \Gamma^{\tau \, (0)}_{r \tau} - \partial_{\tau} \Gamma^{\alpha \, (0)}_{r \alpha} + \Gamma^{\alpha \, (0)}_{r \tau} \Gamma^{\beta \, (2)}_{\alpha \beta} + \Gamma^{\alpha \, (1)}_{r \tau} \Gamma^{\beta \, (1)}_{\alpha \beta} + \Gamma^{\alpha \, (2)}_{r \tau} \Gamma^{\beta \, (0)}_{\alpha \beta} \nonumber \\
&- \Gamma^{\beta \, (0)}_{r \alpha} \Gamma^{\alpha \, (2)}_{\tau \beta} - \Gamma^{\beta \, (1)}_{r \alpha} \Gamma^{\alpha \, (1)}_{\tau \beta} - \Gamma^{\beta \, (2)}_{r \alpha} \Gamma^{\alpha \, (0)}_{\tau \beta} = \frac{a_5}{4 r_c^2} p v^2 \, , \nonumber \\
R_{r i}^{(2)} &= \partial_r \Gamma^{r \, (2)}_{r i} + \partial_j \Gamma^{j \, (1)}_{r i} + \partial_{\tau} \Gamma^{\tau \, (0)}_{r i} - \partial_i \Gamma^{\alpha \, (1)}_{r \alpha} + \Gamma^{\alpha \, (2)}_{r i} \Gamma^{\beta \, (0)}_{\alpha \beta} + \Gamma^{\alpha \, (1)}_{r i} \Gamma^{\beta \, (1)}_{\alpha \beta} + \Gamma^{\alpha \, (0)}_{r i} \Gamma^{\beta \, (2)}_{\alpha \beta} \nonumber \\
&- \Gamma^{\beta \, (2)}_{r \alpha} \Gamma^{\alpha \, (0)}_{i \beta} - \Gamma^{\beta \, (1)}_{r \alpha} \Gamma^{\alpha \, (1)}_{i \beta} - \Gamma^{\beta \, (0)}_{r \alpha} \Gamma^{\alpha \, (2)}_{i \beta} = 0 \, , \nonumber \\
R_{\tau \tau}^{(2)} &= \partial_r \Gamma^{r \, (2)}_{\tau \tau} + \partial_i \Gamma^{i \, (1)}_{\tau \tau} + \partial_{\tau} \Gamma^{\tau \, (0)}_{\tau \tau} - \partial_{\tau} \Gamma^{\alpha \, (0)}_{\tau \alpha} + \Gamma^{\alpha \, (0)}_{\tau \tau} \Gamma^{\beta \, (2)}_{\alpha \beta} + \Gamma^{\alpha \, (1)}_{\tau \tau} \Gamma^{\beta \, (1)}_{\alpha \beta} + \Gamma^{\alpha \, (2)}_{\tau \tau} \Gamma^{\beta \, (0)}_{\alpha \beta} \nonumber \\
&- \Gamma^{\beta \, (0)}_{\tau \alpha} \Gamma^{\alpha \, (2)}_{\tau \beta} - \Gamma^{\beta \, (1)}_{\tau \alpha} \Gamma^{\alpha \, (1)}_{\tau \beta} - \Gamma^{\beta \, (2)}_{\tau \alpha} \Gamma^{\alpha \, (0)}_{\tau \beta} = - \frac{a_5}{4 r_c^2} (r p) v^2 \, , \nonumber \\
R_{\tau i}^{(2)} &= \partial_r \Gamma^{r \, (2)}_{\tau i} + \partial_j \Gamma^{j \, (1)}_{\tau i} + \partial_{\tau} \Gamma^{\tau \, (0)}_{\tau i} - \partial_i \Gamma^{\alpha \, (1)}_{\tau \alpha} + \Gamma^{\alpha \, (0)}_{\tau i} \Gamma^{\beta \, (2)}_{\alpha \beta} + \Gamma^{\alpha \, (1)}_{\tau i} \Gamma^{\beta \, (1)}_{\alpha \beta} + \Gamma^{\alpha \, (2)}_{\tau i} \Gamma^{\beta \, (0)}_{\alpha \beta} \nonumber \\
&- \Gamma^{\beta \, (0)}_{\tau \alpha} \Gamma^{\alpha \, (2)}_{i \beta} - \Gamma^{\beta \, (1)}_{\tau \alpha} \Gamma^{\alpha \, (1)}_{i \beta} -\Gamma^{\beta \, (2)}_{\tau \alpha} \Gamma^{\alpha \, (0)}_{i \beta} = 0 \, , \nonumber \\
R_{i j}^{(2)} &= \partial_r \Gamma^{r \, (2)}_{i j} + \partial_l \Gamma^{l \, (1)}_{i j} + \partial_{\tau} \Gamma^{\tau \, (0)}_{i j} - \partial_j \Gamma^{\alpha \, (1)}_{i \alpha} + \Gamma^{\alpha \, (0)}_{i j} \Gamma^{\beta \, (2)}_{\alpha \beta} + \Gamma^{\alpha \, (1)}_{i j} \Gamma^{\beta \, (1)}_{\alpha \beta} + \Gamma^{\alpha \, (2)}_{i j} \Gamma^{\beta \, (0)}_{\alpha \beta} - \Gamma^{\beta \, (0)}_{i \alpha} \Gamma^{\alpha \, (2)}_{j \beta} \nonumber \\
&- \Gamma^{\beta \, (1)}_{i \alpha} \Gamma^{\alpha \, (1)}_{j \beta} - \Gamma^{\beta \, (2)}_{i \alpha} \Gamma^{\alpha \, (0)}_{j \beta} = \frac{(a_1 - a_2)}{r_c} \partial_{(i} v_{j)} + \frac{(a_4 - a_1^2)}{2 r_c^2} v_i v_j + \frac{1}{2 r_c^2} (a_5 v^2 \delta_{i j} + a_6 r_c \partial_{(i}v_{j)}) \, , \nonumber \\
\end{align}
where $p = \delta_{i i}$ refers to the number of angular dimensions. 


\end{document}